# Integrated ultra-high-performance graphene optical modulator

Elham Heidari[1,†], Hamed Dalir[2,†,*], Farzad Mokhtari Koushyar[1], Behrouz Movahhed Nouri[2], Chandraman Patil[2], Mario Miscuglio[2], Deji Akinwande[1], Volker J. Sorger[2,*]

[1]Microelectronics Research Center, Electrical and Computer Engineering Department, University of Texas at Austin, Austin, Texas 78758, USA
[2]Department of Electrical and Computer Engineering, George Washington University, Washington. DC 20052
†These authors equally contributed to this work
*Corresponding authors: hdalir@gwu.edu

**Abstract:** With the increasing need for large volumes of data processing, transport, and storage, optimizing the trade-off between high-speed and energy consumption in today's optoelectronic devices is getting increasingly difficult. Heterogeneous material integration into Silicon- and Nitride-based photonics has showed high-speed promise, albeit at the expense of millimeter- to centimeter-scale footprints. The hunt for an electro-optic modulator that combines high speed, energy efficiency, and compactness to support high component density on-chip continues. Using a double-layer graphene optical modulator integrated on a Silicon photonics platform, we are able to achieve 60 GHz speed (3 dB roll-off), micrometer compactness, and efficiency of 2.25 fJ/bit in this paper. The electro-optic response is boosted further by a vertical distributed-Bragg-reflector cavity, which reduces the driving voltage by about 40 times while maintaining a sufficient modulation depth (5.2 dB/V). Modulators that are small, efficient, and quick allow high photonic chip density and performance, which is critical for signal processing, sensor platforms, and analog- and neuromorphic photonic processors.

## Introduction

Increased demand for data-handling performance in signal interconnects, optical transceiver technology, and emerging applications such as photonic-based application-specific integrated circuits (ASICS) for analog signal processing or neuromorphic computing and machine-learning acceleration [1] is driving a roadmap for continuous next-generation optoelectronic components [2], including component miniaturization for increased photonic integrated circuit (PIC) density [3].

For example, the bandwidth compression approach used in electrical interconnects is insufficient to meet the rising demand in data processing and telecommunications, owing to capacitive charging and thermal budget constraints. Both challenges may be addressed by photonic interconnects, which decouple bit generation from data transmission and also provide multiplexing parallelization paradigms [4]. Electro-optic modulators are critical components of photonics circuits because they convert analog electrical signals to optical signals by adjusting the effective refractive index (phase, frequency, polarization, or amplitude) of propagation [5,6]. Their performance is dictated by the response time (3dB role-off speed), driving voltage (power consumption), and physical footprint of the device, all of which contribute to the PIC component density and application range. Silicon-based modulators give a mid-range performance, with speeds in the tens of GHz (typically < 50GHz) but are rather bulky (O(~mm)) and need a higher voltage owing to the weak Pockels and Franz-Keldysh effects, but are monolithic to Silicon photonics [7,8]. Germanium-based modulators often have a significant insertion loss and are difficult to integrate monolithically in complementary metal-oxide-semiconductor (CMOS) technology [9,10]. The low material and waveguide losses of lithium niobate enable low driving voltages in exchange for a small footprint (O(~cm)) [11]. As a result, the hunt for an electro-optic modulator that satisfies all critical parameters continues. By using graphene as the modulator material, all device metrics may be addressed; the high mobility enables a 50 Ohm resistance. When paired with strong (beyond unity) optical index modulation at near infrared frequencies, device footprints of less than 10 micrometers provide significant signal modulation (e.g. extinction ratio) and rapid modulation owing to the low RC latency. Current state-of-the-art speeds range between 30-40 GHz [12-14] but fall short of the crucial 50 GHz given by photonic foundries [AIM, AMF, IMEC]. In terms of system integration, graphene research has exhibited CMOS compatibility through wafer-scale integration on silicon [15] and a high material thermal stability provided by its high thermal conductivity. This enables the use of graphene-based photonics not just in cloud-based systems, but also in network-edge devices [5,16]. When these features and capacities are combined, graphene becomes a promising material for light-matter interaction increased high-performance electro-optic devices for PIC integration. We present a graphene-based electro-absorption modulator integrated into silicon photonic waveguides that is capable of switching at 60 GHz, is micrometer compact (6 μm²), and is very efficient (2.25 fJ/bit). The low sub-voltage operation is enabled by (i) packaging a dual-graphene stack in a distributed Bragg reflector (DBR), which increases the path length for light-matter interaction, (ii) Graphene's low resistance of 50 ohm, and (iii) Graphene's natural strong index modulation due to Pauli-blocking at near infrared frequencies (here = 1550nm). The contributions of this demonstration include being the graphene-based modulator in PICs and demonstrating a revolutionary paradigm for extending the path length without compromising the footprint, which increases energy efficiency by 40 times over the traditional design [13].

## Structure and Results

The graphene optical modulator's schematic is shown in Fig. 1 (a). It is based on a Bragg reflector waveguide that is laterally merged with a -Si waveguide. The planar integration of this modulator is advantageous for ultra-dense on-chip optical connection applications. The DBR structure used in this design is composed of alternating layers of high and low refractive index, where the boundary conditions at each layer interface cause partial reflection of electromagnetic waves and the grating period in DBR is designed to cause constructive interference between the reflected beams [17]. Owing to the VCSEL structure's top emission being blocked by a thin film dielectric mirror, light propagates in a zigzag pattern to the modulator cavity [18,19].

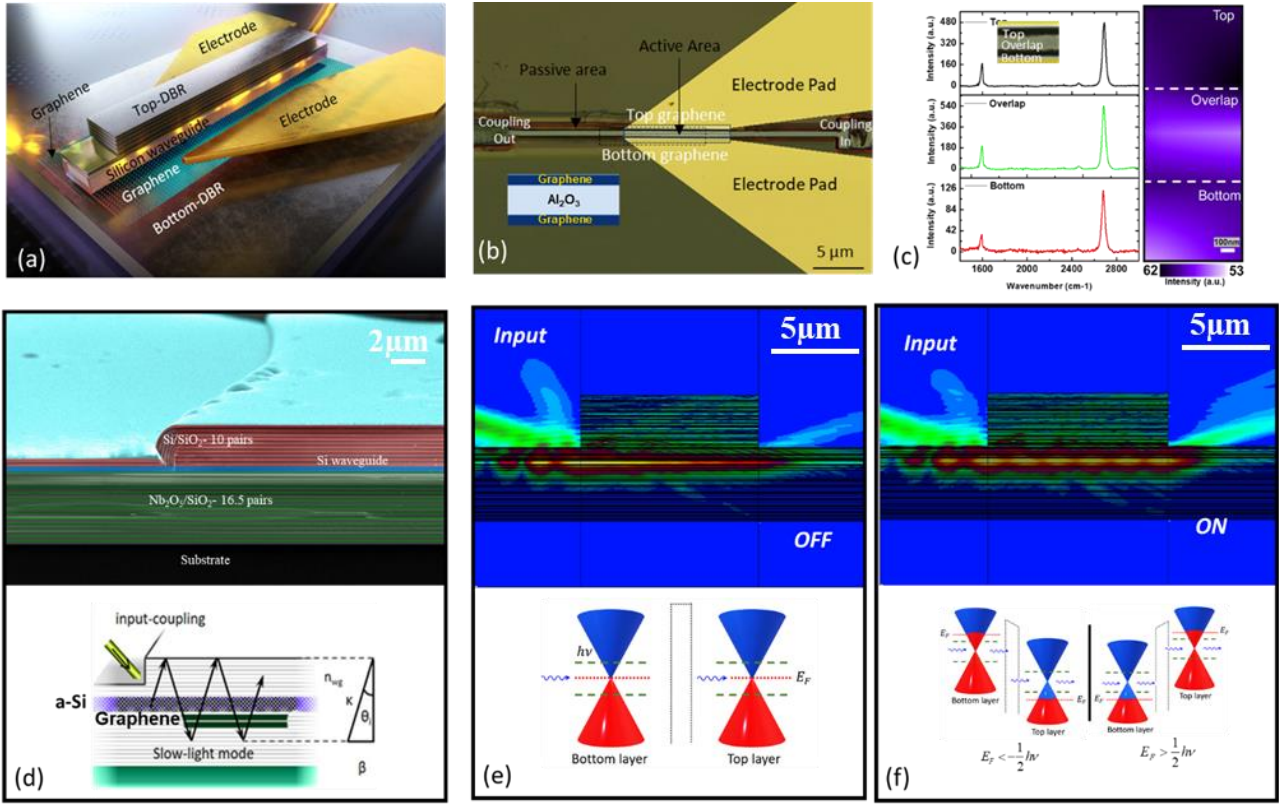

**Fig. 1.** (a) Three-dimensional schematic illustration of the ultra-high performance graphene optical modulator showing two layers of graphene are separated by interlayer of high-κ dielectric ($Al_2O_3$) to form parallel-plate capacitor. Top and bottom DBRs lead to a constructive interference between the reflected beams, (b) Top view microscopic image of the graphene optical modulator showing top view of the waveguiding, input/output couplers and electrode pads, (c) Raman mapping performed across channel showing intensity enhanced at the overlap region (microscope image as inset). (d(i)) Cross-section false-color SEM image of the fabricated graphene optical modulator, (d(ii)) Electric field distributions while excitation using a tilt-coupling scheme (e) Light-emission direction in the modulating cavity. The top emission of the modulator is prohibited by plating a thin-film dielectric reflectors, thereby light propagates to the modulator cavity in a zig-zag shape. EO modulator with un-doped graphene sheets leading Fermi level would be at the Dirac point under the illumination of photons with energy of $\hbar v$. In this case, the transmission would be heavily attenuated, and we called it as OFF state. Graphene's Fermi level distribution under OFF state is shown while the double layer graphene sheets are un-doped. Blue and red colors represent unoccupied and occupied states of Fermi level, respectively. (f) Electric field distribution in EO modulator with graphene's heavily hole or electron doped, which the state is called ON state; whereby inter-band transition is heavily suppressed, and transmission is allowed.

The top view optical microscopic image and cross section false-color scanning electron microscope (SEM) image of the graphene optical modulator are shown in Fig. 1(b) and 1(d(i)), respectively. On an InP substrate, 33 stacked pairs of $Nb_2O_5/SiO_2$ bottom DBRs are constructed. The graphene grown on the copper substrate is transferred to the fabricated bottom DBR to form the core with a planar graphene capacitor using the wet transfer method. Electron beam (e-beam) lithography is used to pattern the electrical contact to each graphene layer. The recommended Pd/Au (10/90 nm) electrode is deposited using an e-beam evaporator. The bottom graphene layer is patterned by e-beam lithography. After that, the undesired graphene is etched by plasma ashing, leaving only the patterned graphene. With e-beam evaporation, a thin coating of Al is deposited on the patterned graphene. A thin (20 nm) $Al_2O_3$ layer is placed on the structure using atomic layer deposition (ALD) for ultra-fast opto-electric applications and to prevent optical mode leakage. The width and thickness of the Si waveguide was designed to be 0.6 μm and 0.445 μm, respectively, to ensure that a single transverse electric (TE) mode is propagated in waveguide. Plasma enhanced chemical vapor deposition (PECVD) was used to deposit an α-Si layer on the graphene and $Al_2O_3$ spacer, which was then e-beam lithographed. Two grating couplers are fabricated, which couples light into and out of the optical waveguide. Finally, 20 layers of $Si/SiO_2$ were deposited with a layer width of $\lambda/4n$, where $\lambda$ and $n$ represents wavelength and refractive index, respectively.

In other words, the Fabry-Perot interferometer is formed by the top and bottom DBRs. A core with a planar graphene capacitor consists of two graphene layers and a high-dielectric material, here Al2O3, interlayer constructed beneath the silicon waveguide to assure the device's high-speed mechanism. The core's effective refractive index is higher than the top and bottom DBR mirrors' effective refractive indices. When a light wave strikes any interface, the angle of incidence exceeds the critical angle for total internal reflection (TIR) at either the bottom or top interfaces, causing the light to be confined and move in a zigzag pattern. The slow group velocity of light induced by light zigzag radiation considerably reduces the size of the optical device, boosting the device's performance, which is shown in Fig. 1(d(ii)).

Variations in the gate-controlled optical absorption in graphene induce changes in the optical modulator's transmission, which is controlled by changing chemical potential (μ, also Fermi level $E_F$) via an external gate field [20]. Electrical gating is used to control the absorption of graphene layers by electrodes connected to each graphene layer, as shown in Fig. 1(e,f) insets at bottom. Graphene's band structure is composed of two bands (valence band (π) and conduction band (π*)) that degenerate at the so-called Dirac point.

The position of the Fermi level can be easily adjusted by changing the accumulation charge because of the low density of states. The Fermi level is at the Dirac point of all graphene layers in an undoped graphene, resulting in low insertion loss (see Fig. 1(e)). Charge transmission would be greatly affected if photons with less than $\hbar v$ energy were used to illuminate. This state is referred to "OFF" state. While Fig. 1(f) represents an integrated double layer graphene hollow core waveguide in the "ON" state, when the graphene sheets are electron-, or hole-doped with emission of photons with the energy of $\hbar v$, and thus the Fermi level increases or decreases, respectively. Transmission between graphene bands is also possible when the Fermi level is changed by tunning the photon's half energy below or above the graphene's Dirac point. However, altering the bias voltage causes the Fermi level to shift below or over the threshold value ($\hbar \omega / 2$), switching the optical inter-band transition ON or OFF.

The input wavelength can be used to modify the incident angle of the "zig-zag" light into the Bragg reflector waveguide, as shown in Eq. (1). Furthermore, the resonance wavelength is determined by the thickness of the α-Si waveguide squeezed between the mirrors (cut off condition: 1565 nm). As $\theta_i$ approaches 90 near the resonance condition, the interaction of light with graphene sheets rises dramatically, enlarging the absorption in the graphene layers.

$$\sin \theta_i = \sqrt{1 - \left(\frac{\kappa_c}{\kappa}\right)^2} = \sqrt{1 - \left(\frac{\lambda}{\lambda_c}\right)^2} \qquad (1)$$

The modulator's performance is critical to the system's overall performance. High modulation speed, large modulation depth, small footprint, and low power consumption are ideal figures of merit (FOM) in the optical modulators. The modulation depth per micrometer of the device was estimated using FIMMWAVE Photon Design Corp.'s film mode matching approach in Fig. 2. At C- and L-band operation, a modulation depth of >4 dB per micron (> 40 dB for a 10 μm long modulator) may be achieved; although insertion loss increases marginally, which is proportional to the increased radiation loss of the DBR mirrors. According to the simulation results, for the on-chip integrated device, the insertion loss is estimated to be 0.7 dB.

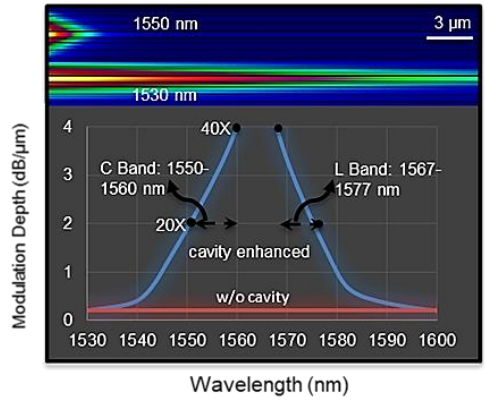

**Fig. 2.** Estimated modulation depth as a function of input-wavelength using film mode matching method of FIMMWAVE Photon Design Corporation. Top image shows the light guide in cavity in 1530nm and 1550 nm wavelengths.

Fig. 3(a) shows the photon transmission through the waveguide when different driving voltages ($V_D$) are applied with wavelengths range from 1.53 to 1.55 μm. The band structure of graphene is made up of two bands that degenerate at Dirac points. One of the properties of Dirac electrons in graphene is that the conductivity of the graphene-based structure changes dramatically when the electrons are confined. The position of the Fermi level in graphene can be easily modified by changing the accumulating charge due to the nature of the monolayer (low density of states). The Fermi level is at the Dirac point in single-layer graphene that is undoped. The transmission would be reduced when photons with an energy of $\hbar v$ (where $\hbar$ and $v$ denote the reduced Plank's constant and light frequency, respectively) were illuminated. Regardless, graphene's Fermi level decreases or increases in either hole- or electron-doped graphene sheets.

As we move from 1530 nm to 1550 nm, the angle of light moves to vertical and light matter interaction enhances so the modulation depth improves. Fig. 3(b) shows the measurement setup schematic for the graphene optical modulator's small signal frequency measurement ($S_{21}$), which is obtained by generating a low power modulating signal (0 dBm) with an Agilent E8361 67 GHz vector network analyzer (VNA) and combining DC voltage bias with the RF signal with a 60 GHz bias tee. The voltage is applied between the top and bottom graphene layers. The light emitted into the modulator by a distributed feedback laser (DFB) at 1550nm. A photodetector (PD, 60 GHz) with a broadband post-amplifier collects modulated light, which is then compared to the original modulating source. The double-layer graphene modulator's dynamic response is shown in Fig. 3(c). The device's useable frequency range extends the optical 3dB (electrical 6dB) bandwidth up to ~60 GHz. The graphene modulator exhibits a 60 GHz small signal radiofrequency (RF) bandwidth, with an RC value restricted by graphene sheet resistance, capacitor size, and contact resistance. The total resistance is expected to be ~50 Ω.

Considering the resistance of the modulator ~ 100 Ω (where 50 Ω comes from the measurement setup system) and C=27 *fF*, we can fairly elaborate the modulation response and power consumption ($E_{bit}$=0.25CV$^2$) of our device. A voltage swing of 0.57 V is necessary to modulate the device at 3dB. This voltage state uses about 2.25 fJ/bit of electricity for the switching operation.

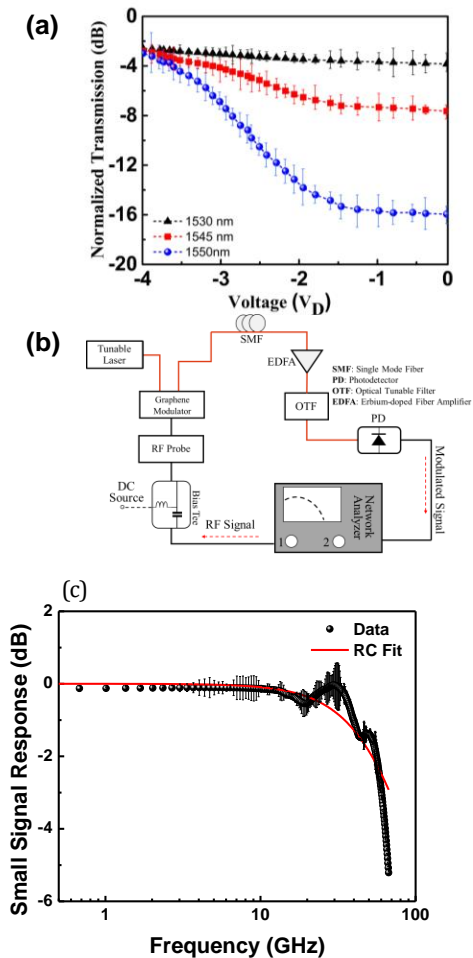

Fig. 3. (a) Electro-optical response of the device at different drive voltage for wavelnegth of 1530nm, 1545nm,1550nm. As we move from 1530 nm to 1550 nm, the angle of light moves to vertical and light matter interaction enhances so the modulation depth improves. The modulation depth of the modulator is estimated to be 5.2 dB/V (b) In-house small-signal RF setup ($S_{21}$: ratio between optical amplitude modulation and RF signal), and (c) Normalized RF response of the double layer graphene optical modulator. The 3 dB cut-off frequency of 60 GHz was obtained.

While charge accumulation is sufficient to elevate the Fermi level with half of the photon's energy above (or below) the Dirac point, the inter-band transition is greatly suppressed, allowing for higher transmission. The strength of the coupling between the evanescent waves and graphene layers is adjusted by modifying chemical potential (also $E_F$) via an external gate field, which causes changes in graphene absorption, resulting in the optical modulator's transmission. Changing the bias voltage causes the fermi level to shift below or above the threshold value, allowing optical interband transitions to be turned on or off.

Conclusion

Here, we demonstrate that by increasing the interaction of light with the graphene layer, by more than 40 times over the standard design, improving light absorption in graphene while reducing the power consumption. The electrical gating used here, adjusts the absorption of graphene layers through electrodes attached to each graphene layer. Because of the direct relationship between applied voltage and absorption in graphene, this device shows capability of sub-volt operation with a sufficient modulation depth (>5 dB). To move the Fermi-level from Dirac to the target point in the Pauli Blocking mechanism, only 2.25 fJ/bit power consumption is required. Due to the graphene's exceptionally high electrical and thermal conductivity, it would also be possible to operate at a fast speed. With the new platform, the device's speed is improved to reach ~60 GHz and beyond.

**Acknowledgment**.
The authors thank Dr. Gernot Pomrenke for his fruitful discussions.